# Front-end Electronics for the CALICE/EUDET Calorimeters

Rémi Cornat[1] on behalf of the CALICE[2] collaboration[*]

1 – Laboratoire Leprince-Ringuet-Ecole Polytechnique/CNRS-IN2P3
Palaiseau – France

2 – http://polywww.in2p3.fr/activites/physique/flc/calice.html

The CALICE collaboration is involved in the design of compact calorimeters featuring a high granularity. The technical developments have to overcome various design issues such as the power dissipation, the integration of the front-end electronics inside the detector and connections between the parts. A huge collaborative work is required to achieve the devices using common tools and designs.

## 1   The CALICE/EUDET calorimeters

The CALICE collaboration contributes to the EUDET project proposing some technical solutions about an Electomagnetic Calorimeter, an Analogue Hadronic Calorimeter and a Digital Hadronic Calorimeter together with a common read-out and data acquisition system[1]. The first prototypes for these detectors have validated the principles of the detection scheme from the point of view of physics performances and with the aim to support the Particle Flow Algorithm Method requiring compactness and granularity. These detectors have been build to perform imaging calorimetry. For example the ECAL features about 10 thousands channels in about one hundredth of a cubic meter. A large set of data have been taken in test beams allowing the linearity, resolution and the calibration to be carefully studied as well as the stability and noise of the electronics channels [2].

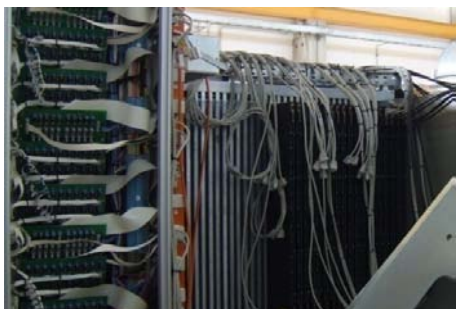

**Figure 1 :** AHCAL first prototype

Following the physics success of a first generation of detectors during years 2004-2008, the second step of prototyping is being reached. The technical and technological feasibility of detectors designs closest to the ILC requirements should now be demonstrated.

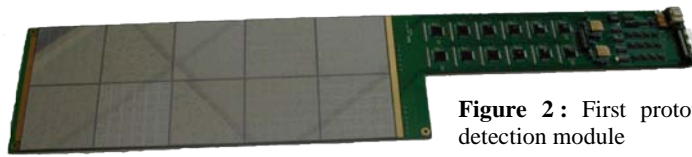

**Figure 2 :** First prototype of ECAL detection module

---

[*] The work described in this paper is partially funded by the EUDET project within the 6[th] Framework program structuring the European research area. (http://www.eudet.org)

## 2 The next generation of front-end electronics

On the contrary to the first prototypes and from the point of view of the electronic system, the next generation of the detectors will feature:

- A front-end electronics fully integrated inside the detectors,
- Front-end integrated chips with analogue to digital conversion and digital read-out layers,
- An optimized power consumption thanks to a power pulsing feature,
- A first level of the data acquisition system together with the detection module.

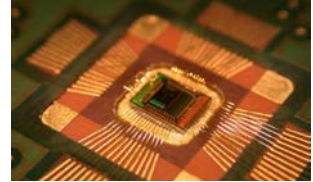

**Figure 3:** The front-end ASIC bounded into the PCB

All the points listed above are major improvements compared to the previous generation. A step by step strategic approach is driven by the CALICE collaboration. The forthcoming prototypes are therefore being designed as technological demonstrators. They are developed taking into account a large set of known constraints given by the ILC environment. The emphasis is put on consequences on the technical aspects rather than on physics performance and data taking.

The various detectors will be held thanks to mechanical structures close to the ones which could be used for an ILC detector. These structures will be partially instrumented: for example, as shown by figure 2, the ECAL will feature only a tower of 400 cm$^2$ cross section together with a long detection layer of 18x150 cm$^2$. Nevertheless it will provide all the necessary hardware to validate the signal integrity, the read-out and data acquisition scheme and the integration of the overall system.

The mechanical structures are based on an assembly of alveoli. The detection modules can be slide into but without both additional space for the front-end chips and the possibility of large scale cooling. It implies all the points listed at the beginning of this section.

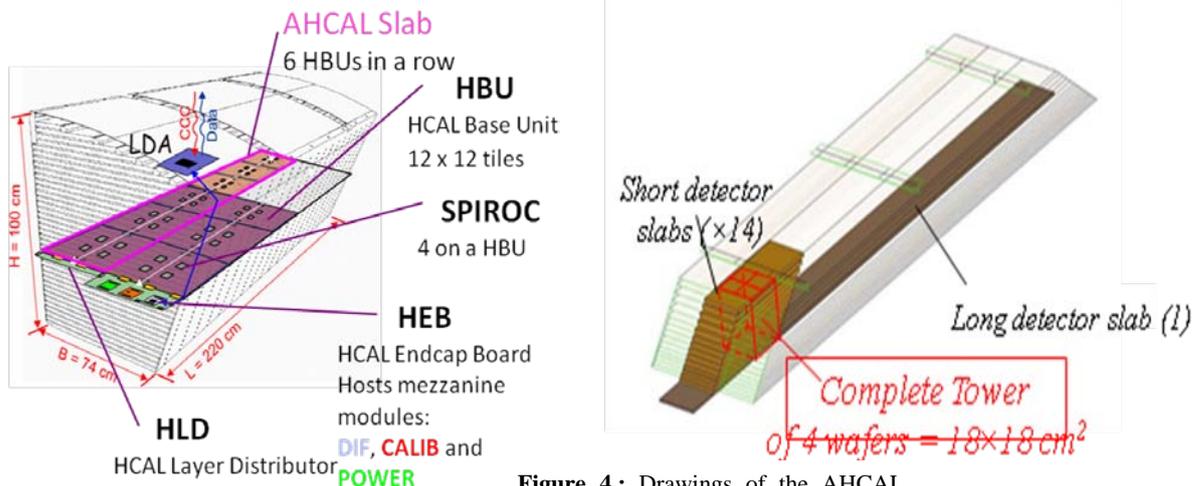

**Figure 4:** Drawings of the AHCAL module (left) and ECAL module (right)

Based on the same building blocks but with some slight change according to the detector type, three flavors of Read Out Chips are designed with a high level of integration. Each of them will include all the features of an LHC front-end board: 64 channels with their own analogue signal processing mutli-gain shapers, an analogue pipe-line, the ADCs, time stamp, and the digital electronics for the read-out, monitoring and slow control.

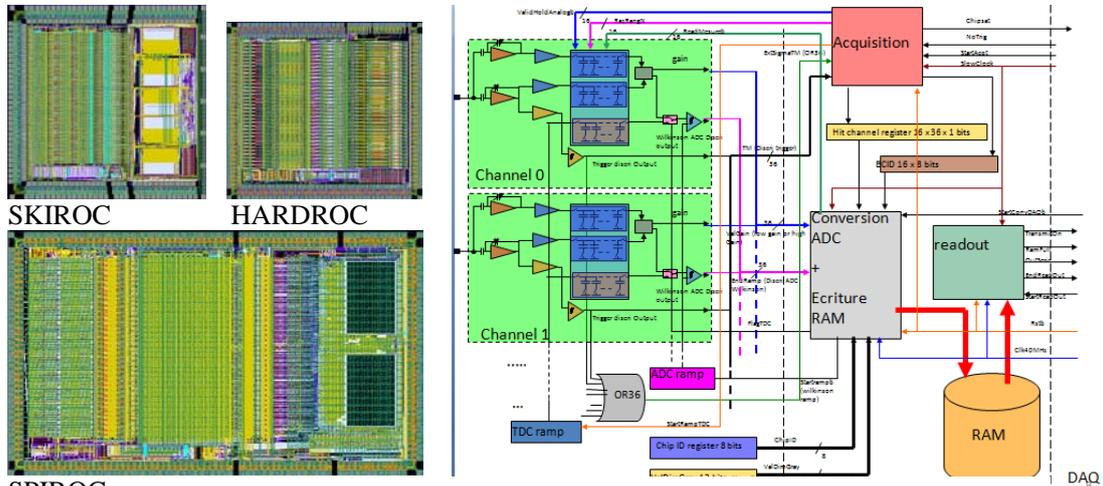

**Figure 5 :** The three ROC circuits (left) and their internal architecture (right)

The chips are designed for low power consumption (25 μW/channel) and for an easy integration as they do not need any external components. The input signal dynamic reach 15 bits and the chips can autonomously trigger at the half MIP level. The amount of data at the output is optimized thanks to an embedded zero-suppress circuitry.

One crucial feature is the power switching capability. The wake-up time is at the order of a few tens of μs (only 2 for the analogue part alone). It makes the ROC chips series fully compliant with the ILC beam structure. Thus the on duty cycle of the chip can be only 0.5% allowing the power budget to be satisfied. For instance 16000 channels of the ECAL should be powered up during one day with only two "AAA" batteries.

A special care is taken about the ADC developments and two options are looked at: a common multiplexed ADC or a low power ADC for each channel. The preferred architecture is pipe-line or cyclic ADCs with multiple bits calculated at each step. Power consumptions of a few mW per ADC at 100% duty cycle are achieved for a noise level below 1 ADC count and a non linearity residual below 3 ADC counts for the whole dynamics.

## 3 The Read-out Electronics

The read-out electronics is based on successive levels of data concentrator linked with increasing speed serial protocols ending with Giga-Ethernet on optical links. A complete description can be found here:[3]. The first level of the data acquisition is the Detector InterFace (DIF). The DIF is located close to the detection planes between the structural modules of the detectors. Only a small space is allocated (less than 20 cm$^3$) but a large set of functionalities have to be featured: read-out and data buffering of about 100 ROC chips, fast signaling and slow control of the ROC chips, interface to the higher level of the DAQ system, power controlling, monitoring of temperature and power supplies.

In addition the DIF faces the signal integrity issues due to the bus topology used for some control signals like the clocks. The switching of signals at a few tens of MHz on up to 2 m long PCB traces with a high distributed capacitive load and linear resistance must be ensured. In addition the power on/off cycles have to be properly controlled to avoid any voltage overrange or instability that could disturb the front-end electronics.

First prototypes of the DIF boards are FPGA based and include an USB link for debugging purpose, serial interfaces to the DAQ system, embedded memory and connectors to the detector modules.

Two other aspects are also critical: the interconnections which have to be reliable and compact (located within the PCB thickness, at the edges) and the cooling of the detector.
For this last point a copper sheet is included with the detector module and is connected to an external cooling pipe. A proper management of the power enable signal for each element of the electronics is performed. Simulations are giving a 10 degrees temperature drop between two ends of a module.

## 4 Conclusion

The next generation of CALICE/EUDET prototypes is planned for the end of the year 2009 with major technological improvements compared to the previous development step. Seen as an engineering demonstrator, integration, power issues, signal integrity and thermal dissipation will be particularly studied. Nevertheless the detectors will be fully functional for physics and data taking. With detectors featuring a few tens of hundreds channels, the granularity as well as the compactness will be improved by a factor 2 to 4 thus impacting on physics performances (particle flow algorithms).